\documentclass[a4paper,11pt]{article}
\pdfoutput=1
\usepackage{jheppub}

\usepackage[utf8]{inputenc}

\newcommand{\Trh}{T_\text{rh}}

\newcommand{\nfi}{n_\text{fi}}
\newcommand{\Tfi}{T_\text{fi}}
\newcommand{\Tp}{{T'}}
\newcommand{\Tpfi}{{T'_\text{fi}}}
\newcommand{\afo}{a_\text{fo}}
\newcommand{\Tfo}{T_\text{fo}}
\newcommand{\Tpfo}{{T'_\text{fo}}}
\newcommand{\xp}{{x'}}
\newcommand{\xpfo}{{x'_\text{fo}}}
\newcommand{\gs}{g_\star}
\newcommand{\gss}{g_{\star s}}
\newcommand{\Br}{\text{Br}}
\newcommand{\Cr}{\mathcal{C}_\rho}
\newcommand{\Cn}{\mathcal{C}_n}

\title{Boosting Freeze-in through Thermalization}

\author{Nicolás Bernal}

\affiliation{Centro de Investigaciones, Universidad Antonio Nariño\\
Carrera 3 Este \# 47A-15, Bogotá, Colombia}

\emailAdd{nicolas.bernal@uan.edu.co}

\abstract{
If the interaction rates between the visible and the dark sectors were never strong enough, the observed dark matter relic abundance could have been produced in the early Universe by non-thermal processes.
This is what occurs in the so-called freeze-in mechanism.
In the simplest version of the freeze-in paradigm, after dark matter is produced from the standard model thermal bath, its abundance is frozen and remains constant.
However, thermalization and number-changing processes in the dark sector can have strong impacts, in particular enhancing the dark matter relic abundance by several orders of magnitude.
Here we show that this enhancement can be computed from general arguments as the conservation of energy and entropy, independently from the underlying particle physics details of the dark sector.
We also note that this result is quite general, and applies to FIMP production independently of being UV- or IR-dominated.
}

\begin{document}
\begin{flushright}
    PI/UAN-2020-671FT
\end{flushright}

\maketitle

\section{Introduction}

The existence of a dark matter (DM) component has been firmly established by astrophysical and cosmological observations, although its fundamental nature remains elusive~\cite{Aghanim:2018eyx}.
Up to now, the only evidence about the existence of such dark component is via its gravitational effects with the standard model (SM).
For a long time, weakly interacting massive particles (WIMPs) have been among the best-motivated DM candidates.
However, the increasingly strong observational constraints on DM are motivating the quest of particle DM models beyond the WIMP paradigm~\cite{Arcadi:2017kky}.

A simple alternative to the WIMP mechanism consists in relaxing the assumption that DM is a thermal relic produced by the freeze-out mechanism.
In fact, the observed DM abundance may also be generated out of equilibrium by the so-called freeze-in mechanism~\cite{McDonald:2001vt, Choi:2005vq, Kusenko:2006rh, Petraki:2007gq, Hall:2009bx, Elahi:2014fsa}.
In that case DM is a feebly interacting massive particle (FIMP).
For a recent review of FIMP DM models and observational constraints see ref.~\cite{Bernal:2017kxu}.

For freeze-in to take place, the DM production rate has to be very suppressed, and much smaller than the Hubble expansion rate, in order to avoid chemical equilibrium with the SM.
One possibility for having small interaction rates is by assuming renormalizable processes connecting the two sectors, and a product of mediator couplings to the SM and to DM in the ballpark of $\mathcal{O}(10^{-11})$.
In that case, dubbed IR freeze-in, the bulk of the DM is typically produced when the SM temperature is of the order of the mediator mass.
Another possibility for having small production rates is by taking non-renormalizable interactions, suppressed by a large dimensional quantity which is parametrically the mass scale of the mediator.
That is the case for the UV freeze-in~\cite{Elahi:2014fsa}, where DM is mainly produced at the highest temperature reached by the SM thermal bath.
This can be the reheating temperature in the case of an instantaneous and complete inflaton decay, but can also be much larger if that approximation is not used for reheating~\cite{Giudice:2000ex}.%
\footnote{Its is interesting to note the recent intensive effort on exploring the effects on the DM produced by UV freeze-in due to more realistic pictures for reheating~\cite{Garcia:2017tuj, Chen:2017kvz, Bernal:2018qlk, Bhattacharyya:2018evo, Garcia:2018wtq, Chowdhury:2018tzw, Kaneta:2019zgw, Harigaya:2019tzu, Banerjee:2019asa, Bernal:2019mhf, Chanda:2019xyl, Baules:2019zwk, Dutra:2019xet, Dutra:2019nhh, Mahanta:2019sfo, Cosme:2020mck, Garcia:2020eof, Bernal:2020bfj, Bernal:2020qyu}.}

However, independently from the fact of being IR- or UV-dominated, the bulk of FIMP DM is produced in a short period of time, after which the two sectors decouple.
Additionally, the DM abundance remains constant provided that there are no sizable interactions within the dark sector.
The dynamics in the dark sector could be nevertheless more complicated, featuring for example $N$-to-$N'$ number-changing processes, where $N$ DM particles annihilate into $N'$ of them (with $N>N'\geq 2$).
The dominant $N$-to-$N'$ interactions typically correspond to 3-to-2 (see e.g. refs.~\cite{Carlson:1992fn, Hochberg:2014dra, Bernal:2015bla, Bernal:2015lbl, Bernal:2015ova, Pappadopulo:2016pkp, Farina:2016llk, Choi:2017mkk, Chu:2017msm}), but are forbidden in the most common models where the DM stability is guaranteed by a $\mathbb{Z}_2$ symmetry.
In that case, unavoidable 4-to-2 annihilations~\cite{Bernal:2015xba, Heikinheimo:2016yds, Bernal:2017mqb, Heikinheimo:2017ofk, Bernal:2018ins, Bernal:2018hjm} could dominate.
If number-changing processes in the dark sector reach equilibrium, DM forms  a thermal bath with a temperature in general different from the one of the SM.
More importantly, these number-changing processes have a strong impact on DM, increasing by several orders of magnitude its relic abundance.

In this work we investigate the impact of thermalization of the dark sector on the DM abundance produced by the FIMP mechanism.
In particular we highlight that following general entropy conservation considerations, after freeze-in production the evolution of the DM density does not depend on the given particle physics details of the dark sector, but rather on the moment at which the number-changing interactions in the dark sector decouple.
We also note that this framework is quite general, and applies to FIMP production independently from being UV- or IR-dominated.

\section{FIMP Dark Matter Production}
\label{sec:DM}

If the interaction rates between the visible and the dark sectors were never strong enough, the observed DM relic abundance could still have been produced in the early Universe by non-thermal processes.
This is what occurs in the so-called freeze-in mechanism.
In the simplest version of the freeze-in paradigm, after DM is produced from the SM thermal bath, its abundance is frozen and remains constant.
However, thermalization and number-changing processes in the dark sector have a strong impact on the DM relic abundance, as it will be seen hereafter.

\subsection{Without Thermalization}

In the freeze-in paradigm the bulk of the DM is generated when the SM bath has a temperature $T=\Tfi$, where $\Tfi$ is typically the reheat temperature $\Trh$ in the case of a UV production, or the mediator mass in the case where the production is IR-dominated.
To track the evolution of the DM number density $n$, it is convenient to define the DM yield $Y$ as the ratio $n$ over the SM entropy density $s$, with
\begin{equation}
    s(T)=\frac{2\pi^2}{45}\gss(T)\,T^3,
\end{equation}
where $\gss$ corresponds to the effective number of degrees of freedom contributing to the SM entropy~\cite{Drees:2015exa}.
The asymptotic value at $T\ll\Tfi$ of the DM yield in the case where no interaction within the dark sector is simply
\begin{equation}
    Y_0^\text{w/o} \simeq Y^\text{w/o}(\Tfi)
    =\frac{\nfi}{s(\Tfi)}\,,
\end{equation}
where $\nfi$ corresponds to the original DM number density produced by freeze-in, strongly depending on the details of the portal connecting the the dark to the visible sector.

A typical example of UV freeze-in corresponds to the DM production via the 2-body decay of the inflaton.
In that case the DM number density is mainly produced at $T\simeq\Trh$ and reads
\begin{equation}\label{eq:nfi}
    \nfi=n(\Trh) = 
    \frac{\pi^2}{15}\gs(\Trh)\,\frac{\Trh^4}{m_\phi}\,\Br\,,
\end{equation}
where $\gs$ corresponds to the effective number of relativistic degrees of freedom contributing to the SM energy density~\cite{Drees:2015exa}, $m_\phi$ is the inflaton mass and $\Br$ the branching fraction for the decay of the inflaton decaying into two DM particles.
The asymptotic value of the DM yield is therefore
\begin{equation}\label{eq:Ywoinflaton}
    Y_0^\text{w/o} \simeq  
    \frac32\frac{\gs(\Trh)}{\gss(\Trh)}\,\frac{\Trh}{m_\phi}\,\Br\,.
\end{equation}
In Starobinsky inflation~\cite{Starobinsky:1980te} typical values are $m_\phi\simeq 3\times 10^{13}$~GeV and $\Trh\simeq 10^8$~GeV, which implies that in order to reproduce the observed DM abundance the branching fraction to DM states has to be
\begin{equation}\label{eq:brwo}
    \Br\simeq 5\times 10^{-8}\left(\frac{1~\text{TeV}}{m}\right),
\end{equation}
with $m$ being the DM mass.
We emphasize that this result ignores possible thermalization and number-changing processes in the dark sector.

\subsection{With Thermalization}

Even if it is typically neglected, in freeze-in scenarios thermalization within the dark sector can occur, inducing a strong impact on the DM abundance~\cite{Chu:2013jja, Bernal:2015ova, Bernal:2015xba, Bernal:2017mqb, Falkowski:2017uya, Herms:2018ajr, Heeba:2018wtf, Mondino:2020lsc}.
Here we show that using simple and general assumptions, the role of thermalization and DM number-changing interactions within the dark sector can be estimated in a model independent way.

The initial DM energy density produced by the freeze-in mechanism can be estimated as
\begin{equation}
    \rho_\text{fi}\simeq \nfi\,\Tfi\,.
\end{equation}
If there exist strong interactions within the dark sector DM rapidly thermalizes, its distribution being characterized by a temperature $\Tp$, in general different from $T$.
Assuming an instantaneous thermalization process, the temperature $\Tpfi$ in the dark sector just after thermalization is
\begin{equation}\label{eq:Tpfi}
    \Tpfi^4\simeq \frac{30}{\pi^2\,\Cr\,g}\,\nfi\,\Tfi\,,
\end{equation}
where $\Cr=1$ (bosonic DM) or $7/8$ (fermionic DM), and $g$ corresponds to the DM degrees of freedom.
In eq.~\eqref{eq:Tpfi} the fact that the DM energy density $\rho$ is given by
\begin{equation}
    \rho(\Tp)\simeq \Cr\frac{\pi^2}{30}\,g\,\Tp^4\,,
\end{equation}
was used.
The DM number density just after thermalization is therefore
\begin{equation}
    n(\Tpfi) = \Cn\frac{\zeta(3)}{\pi^2}\,g\,\Tpfi^3
    \simeq \frac{\Cn\,\zeta(3)}{\pi^\frac72}\,g^\frac14\left[\frac{30}{\Cr}\,\nfi\,\Tfi\right]^\frac34,
\end{equation}
where $\Cn=1$ or $3/4$ for bosonic or fermionic DM, respectively.
Number-changing interactions in the dark sector increase the DM number density from $\nfi$ to $n(\Tpfi)$, at the cost of decreasing the average energy per particle.
The dark sector rapidly cools down until chemical equilibrium is reached.

The evolution of the SM and DM temperatures is dictated by entropy conservation~\cite{Carlson:1992fn}.
In fact, entropies are separately conserved after the moment when the two sectors kinematically decouple from each other, i.e. for $T<\Tfi$.
On the one hand, up to variations in $\gss$ the SM temperature scales like $T(a)\propto 1/a$, where $a$ corresponds to the scale factor.
On the other hand, DM temperature also scales like $\Tp(a)= \Tpfi/a$ as long as DM is ultra-relativistic, i.e. $\Tp\gg m$.
However, it decreases slower when it becomes non-relativistic and it is still chemically coupled, i.e. between $m\gg\Tp\gg\Tpfo$, where $\Tp=\Tpfo$ is the temperature at which the number-changing interactions in the dark sector freeze-out.
In this regime
\begin{equation}
    \Tp(a)=2\,m\,W_0^{-1}\left[\frac{2025}{16\pi^7\,\Cr^2}\left(\frac{m}{\Tpfi}\,a\right)^6\right],
\end{equation}
where $W_0$ corresponds to the 0-branch of the Lambert function.
There is therefore a {\it relative} increase of $\Tp$ compared to $T$.
Finally, for $\Tp\ll\Tpfo$, DM is non-relativistic and out of chemical equilibrium, implying that $\Tp(a)= \Tpfo\,(\afo/a)^2$, where $\afo$ is the scale factor when $\Tp=\Tpfo$ and is given by
\begin{equation}
    \afo^3=\frac{4\sqrt{2\pi^7}\,\Cr}{45}\frac{\Tpfi^3}{\sqrt{\Tpfo\,m^5}}\,e^\frac{m}{\Tpfo}.
\end{equation}

\begin{figure}[t!]
	\begin{center}
		\includegraphics[height=0.33\textwidth]{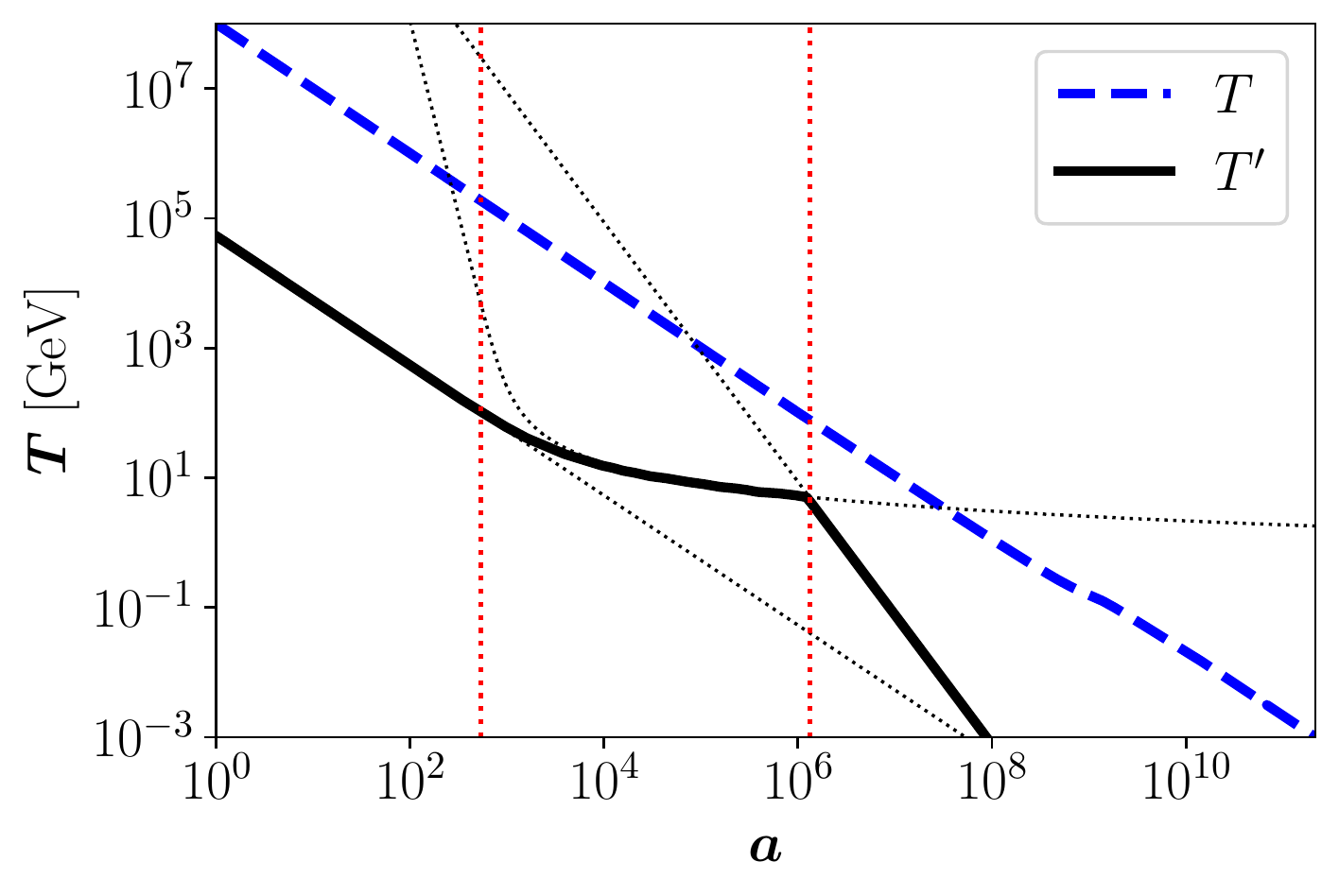}
		\includegraphics[height=0.33\textwidth]{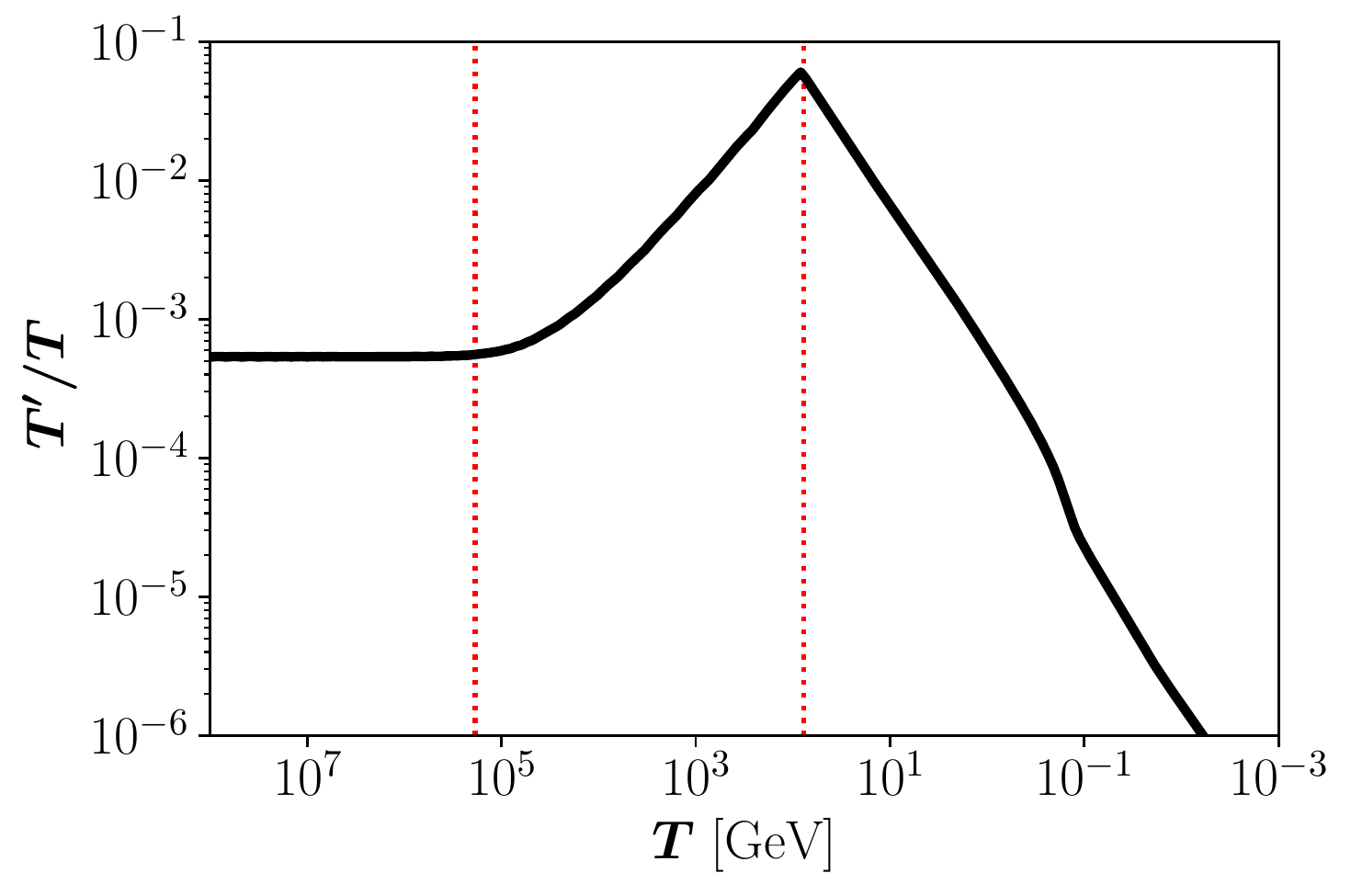}
        \caption{Example of the evolution of the SM temperature $T$ and DM temperature $\Tp$, assuming $m=100$~GeV and $\xpfo=20$.
        The vertical red dotted lines represent $\Tp=m$ (left) and $\Tp=\Tpfo$ (right).
        The black dotted curves correspond to analytical estimations for $\Tp$.
		}
		\label{fig:Temp}
	\end{center}
\end{figure}
The left panel of fig.~\ref{fig:Temp} shows an example of the evolution of the SM temperature $T$ (blue dashed line) and DM temperature $\Tp$ (black solid line) as a function of the scale factor $a$, assuming $m=100$~GeV and $\xpfo\equiv m/\Tpfo=20$.
For the initial DM number density $\nfi$ we have used eq.~\eqref{eq:nfi} together with $\Trh=10^8$~GeV, $m_\phi=3\times 10^{13}$~GeV and $\Br=10^{-10}$.
The vertical red dotted lines representing $\Tp=m$ (left) and $\Tp=\Tpfo$ (right) have been added for reference.
Additionally, the black dotted curves correspond to the analytical estimations for the evolution of $\Tp$, in the three regimes previously described.
Finally, the right panel of fig.~\ref{fig:Temp} shows the ratio of temperatures $\Tp/T$ as a function of $T$, for the same benchmark point.
The small kink near $T\sim 100$~MeV corresponds to the QCD crossover.

The asymptotic value at $\Tp\ll\Tpfo$ of the DM yield in the case with sizable self-interactions within the dark sector depends on the moment when the DM number-changing processes decouple.
In the case $\xpfo\ll 1$, these interactions freeze-out when the DM is still ultra-relativistic, and therefore the DM yield reads
\begin{equation}
    Y^\text{w/}_0=\frac{n(\Tpfi)}{s(\Tfi)}
    \simeq \frac{45\,\Cn\,\zeta(3)}{2\pi^\frac{11}{2}}\frac{g^\frac14}{\gss(\Tfi)}\left[\frac{30}{\Cr}\,\frac{\nfi}{\Tfi^3}\right]^\frac34.
\end{equation}
However, in the opposite case where $\xpfo\gg 1$, the dark freeze-out occurs when DM is non-relativistic, and the DM yield is instead
\begin{equation}
    Y^\text{w/}_0\simeq\frac{n(\Tpfo)}{s(\Tfo)}\simeq \frac{45}{2\pi^2}\frac{g}{\gss(\Tfo)}\left(\frac{m\,\Tpfo}{2\pi}\right)^\frac32\frac{1}{\Tfo^3}\,e^{-\frac{m}{\Tpfo}}\simeq \frac{8}{\pi^\frac32}\frac{g^\frac14}{\gss(\Tfi)}\frac{\Tpfo}{m}\left[\frac{15}{7}\frac{\nfi}{\Tfi^3}\right]^\frac34.
\end{equation}

\begin{figure}[t!]
	\begin{center}
		\includegraphics[height=0.33\textwidth]{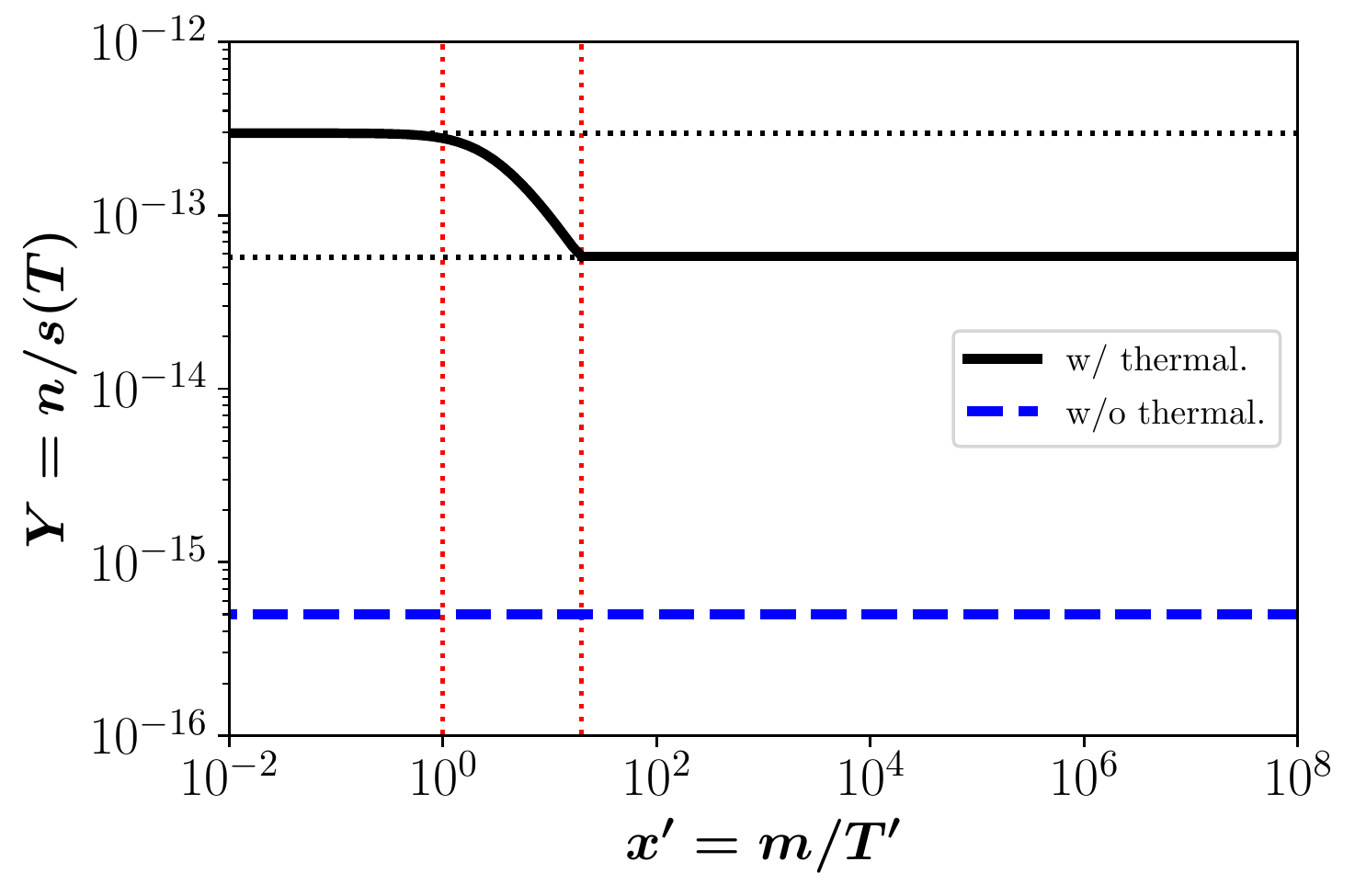}
		\caption{Example of the evolution of the DM yield with (black solid line) and without (blue dashed line) thermalization in the dark sector, assuming $m=100$~GeV and $\xpfo=20$.
        The vertical red dotted lines represent $\Tp=m$ (left) and $\Tp=\Tpfo$ (right), whereas the horizontal black dotted lines correspond to the analytical estimations for the DM yield.
        }
		\label{fig:Yield}
	\end{center}
\end{figure}
Figure~\ref{fig:Yield} presents an example of the evolution of the DM yield with (black solid line) and without (blue dashed line) thermalization within the dark sector, for the same benchmark used in fig.~\ref{fig:Temp}, i.e. $m=100$~GeV and $\xpfo=20$.
The vertical red dotted lines represent $\xp=1$ (left) and $\xp=\xpfo$ (right).
The horizontal black dotted lines correspond to the analytical estimations for the DM yield assuming that the freeze-out occurs when DM is ultra-relativistic (upper line) or non-relativistic (lower line).
The figure shows an enhancement of few orders of magnitude in the produced DM abundance reached comparing the cases with and without thermalization.
This enhancement is maximized if the number-changing interactions within the dark sector decouple when the DM is ultra-relativistic.
In the opposite case, if the interactions freeze-out when DM is non-relativistic, the enhancement is reduced, but only by a factor $1/\xpfo$ and not $e^{-\xpfo}$, due to the close-to-exponential increase of $\Tp$ with respect to $T$ near $\Tp=\Tpfo$.

A measure of the impact of the thermalization on the DM yield can be estimated by defining a boost factor $B$ which is the ratio of the DM abundance taking into account the case with relative to the case without thermalization in the dark sector:
\begin{equation}\label{eq:boost}
    B\equiv\frac{Y^\text{w/}_0}{Y^\text{w/o}_0}\simeq\left(\frac{8}{27}\frac{g}{\gss(\Tfi)}\frac{1}{Y_0^\text{w/o}}\right)^\frac14\times
    \begin{cases}
        \frac{45\,\zeta(3)}{2^{1/4}\,\pi^4}\frac{\Cn}{\Cr^{3/4}}&\qquad\text{ for } \xpfo\ll 1,\\[8pt]
        \frac{8}{7^{3/4}}\frac{1}{\xpfo}&\qquad\text{ for } \xpfo\gg 1.\\
    \end{cases}
\end{equation}
This boost factor is a clean way to characterize the enhancement since many of the other factors fall out.
In particular, we highlight that $B$ mainly depends on $Y_0^\text{w/o}$ and $\xpfo$,%
\footnote{There is also a marginal dependence on $\Tfi$ if it is bellow the electroweak scale, and the spin of the DM.}
but not on $m$ or the specific number-changing processes that brought the dark sector into chemical equilibrium.
We emphasize that the computation of the boost factor is independent on the details of the underlying particle physics details of the dark sector.
In fact, we have only assumed an instantaneous thermalization in the dark sector, and that DM number-changing interactions in the dark sector reached chemical equilibrium.
Both conditions can be naturally fulfilled provided that DM features sizable self-interactions.

Finally, and following the example of the UV frozen-in DM produced by the decay of the inflaton introduced in the previous section, eqs.~\eqref{eq:Ywoinflaton} and~\eqref{eq:boost} imply that the branching ratio of the decay for the inflaton into a couple of DM particles has to be
\begin{equation}
    \Br \simeq
    10^{-10}\left(\frac{1~\text{TeV}}{m}\right)^\frac43\times
    \begin{cases}
        1&\qquad\text{ for } \xpfo\ll 1,\\[8pt]
        0.2\,\xpfo^\frac43&\qquad\text{ for } \xpfo\gg 1,\\
    \end{cases}
\end{equation}
in order to reproduce the observed DM abundance, for real scalar DM.
The thermalization and number-changing processes in the dark sector enhance the DM abundance, decreasing the required branching fraction of the inflaton into DM states.
We note additionally, that $\Br$ presents a stronger mass dependence.

\section{Conclusions}
\label{sec:end}

Dark matter has been typically assumed to be a thermal relic produced via the WIMP mechanism.
However, if the interaction rates between the visible and the dark sectors were never strong enough, the observed DM relic abundance could still have been produced in the early Universe by non-thermal processes.
This is what occurs in the so-called freeze-in mechanism.

In the simplest version of the freeze-in paradigm, after DM is produced from the SM thermal bath, its abundance is frozen and remains constant.
Nevertheless, {\it thermalization and number-changing processes in the dark sector can have strong impacts, in particular enhancing the DM relic abundance by several orders of magnitude.}
Here we have shown that the boost can be computed from general arguments as the conservation of energy and entropy, independently from the underlying particle physics details of the dark sector.
We also note that this result is quite general, and applies to FIMP production regardless of being UV- or IR-dominated.

Before concluding, we note that thermalization and number-changing interactions naturally appear in scenarios where DM features sizable self-interactions.
Those DM self-interactions could play a role in the solution of the so-called `core vs. cusp problem'~\cite{Flores:1994gz, Moore:1994yx, Oh:2010mc, Walker:2011zu} and `too-big-to-fail problem'~\cite{BoylanKolchin:2011de, BoylanKolchin:2011dk, Garrison-Kimmel:2014vqa, Papastergis:2014aba} arising at small scales.
For this to be the case, the required self-scattering cross section over DM mass needs to be of the order of 0.1--2~cm$^2$/g at the scale of dwarf galaxies~\cite{Kaplinghat:2015aga, Fry:2015rta}, and smaller than $1.25$~cm$^2$/g at the scale of galaxy clusters~\cite{Randall:2007ph}.
Further constraints on this scenario come from the number of relativistic degrees of freedom $N_\text{eff}$ at the CMB and BBN epochs, which constrain DM to be heavier than few keV~\cite{Cyburt:2015mya, Bernal:2017mqb, Aghanim:2018eyx}.
Finally, let us note that self-scattering is also relevant in the high-redshift Universe as it controls the free-streaming length ($\lambda_\text{fs}$) of DM particles, which in turn determines the smallest DM objects that can be formed from primordial perturbations~\cite{Profumo:2006bv}.
The strongest observationally inferred limit on $\lambda_\text{fs}$ is derived from the matter power spectrum suppression induced by DM free-streaming and comparing Lyman-$\alpha$ observations and cosmological hydrodynamical simulations.
The current limit $\lambda_\text{fs}\lesssim 100$~kpc, which in this kind of models translates into a lower bound on the DM mass of few keV~\cite{Bernal:2017mqb}.

\section*{Acknowledgments}
NB thanks Xiaoyong Chu, Johannes Herms and Hardi Veermäe for valuable discussions.
NB is partially supported by Universidad Antonio Nariño grants 2018204, 2019101 and 2019248, and by Spanish MINECO under Grant FPA2017-84543-P. 
This work was supported by the European Union's Horizon 2020 research and innovation program under the Marie Sk\l{}odowska-Curie grant agreements 674896 and 690575.

\bibliographystyle{JHEP}
\bibliography{biblio}

\end{document}